\documentclass[aps,pra,reprint,superscriptaddress,showpacs]{revtex4-1}

\usepackage{graphicx}
\usepackage{amsmath}
\usepackage{color}
\usepackage{array}

\usepackage{ulem}

\begin{document}

\title{
Soft landing of metal clusters on graphite: a molecular dynamics study
}

\author{Alexey V. Verkhovtsev}
\email[]{verkhovtsev@mbnexplorer.com}
\altaffiliation{On leave from Ioffe Institute, Politekhnicheskaya 26, 194021 St. Petersburg, Russia}
\affiliation{MBN Research Center, Altenh\"oferallee 3, 60438 Frankfurt am Main, Germany}

\author{Yury Erofeev}
\affiliation{MBN Research Center, Altenh\"oferallee 3, 60438 Frankfurt am Main, Germany}
\affiliation{Department of Physics, Utrecht University, Heidelberglaan 8, 3584 CS Utrecht, The Netherlands}

\author{Andrey V. Solov'yov}
\altaffiliation{On leave from Ioffe Institute, Politekhnicheskaya 26, 194021 St. Petersburg, Russia}
\affiliation{MBN Research Center, Altenh\"oferallee 3, 60438 Frankfurt am Main, Germany}


\begin{abstract}
Structure and stability of 3 nm size Ag$_{887}$, Au$_{887}$ and Ti$_{787}$ clusters
deposited on graphite under soft landing conditions ($\sim 10^{-3}- 10^0$~eV per atom)
are studied by means of molecular dynamics simulations.
Parameters for the cluster--surface interaction are derived from complementary \textit{ab initio} calculations.
We found that the shape of clusters on the surface is governed by their elemental composition
and depends also on the initial cluster structure and landing conditions.
At deposition energies below 0.1~eV/atom, the Ag$_{887}$ cluster acquires an ellipsoidal shape,
while Au$_{887}$ and Ti$_{787}$ clusters transform into oblate and prolate truncated spheroids, respectively,
due to stronger adhesion to graphite.
The clusters flatten over the surface and eventually disintegrate as the deposition energy increases.
Simulation results reveal that Ag$_{887}$ and Au$_{887}$ fragment at about $0.75 - 1.0$~eV/atom
whereas higher energy of about 3 eV/atom is required for the fragmentation of Ti$_{787}$.
The contact angle, contact radius and height of the clusters as functions of deposition energy
are determined from fitting the positions of cluster surface atoms with a surface equation.
The dependence of these parameters on internal energy of the clusters is also analyzed.
\end{abstract}

\maketitle

\section{Introduction}

The interaction of atomic clusters and nanoparticles with surfaces has been a widely studied topic
in cluster science over the past several decades \cite{Jensen_1999_RMP.71.1695, Meiwes-Broer_book, ISACC_LatestAdv_2008}.
The dynamics of metal clusters and carbon fullerenes deposited onto different surfaces
(mainly, metal surfaces, graphite and silicon oxide) was explored
both experimentally \cite{Bromann_1996_Science.274.956, Carroll_1996_JPCM.8.L617,
Hillenkamp_2002_JCP.116.6764, Kaplan_2009_PRB.79.233405}
and computationally by means of molecular dynamics (MD) simulations
\cite{Cheng_1994_JPC.98.3527, Thaler_2014_JCP.140.044326, Bernstein_2016_JCP.145.044303}.
The strong interest in the deposition of mass-selected clusters on surfaces has been
motivated by both fundamental research and technological applications.

From a fundamental physics viewpoint, an important question is how
structural, electronic, magnetic and optical properties of deposited clusters
change with respect to free counterparts.
A variety of phenomena emerge also when atomic clusters are brought in contact with a surface.
Examples include fragmentation of clusters and implantation of cluster atoms
into the substrate \cite{Pratontep_2003_PRL.90.055503},
penetration of energetic clusters through a substrate and surface sputtering
\cite{Smith_1993_ProcRoyalSocA.441.495, Plant_2014_Nanoscale.6.1258},
irradiation-induced structural rearrangements of deposited clusters \cite{Wang_2012_PRL.108.245502},
cluster diffusion and aggregation into islands \cite{Bardotti_1995_PRL.74.4694, Alayan_2007_PRB.76.075424},
super-diffusion and L\'{e}vy flights \cite{Luedtke_1999_PRL.82.3835},
as well as the formation and fragmentation of fractal nanostructures
\cite{Brechignac_2002_PRL.88.196103, Lando_2007_EPJD.43.151, Solovyov_2014_PSSB.251.609}.

From a technological viewpoint, an understanding of the cluster--surface interaction
is crucial for the controllable production of novel materials such as
thin films and nanostructured surfaces \cite{Palmer_2003_NatureMater.2.443},
supported nanocatalysts \cite{Vajda_2015_ACSCatal.5.7152} as well as
nanoscale components for electronic devices \cite{ClusterBeamDeposition_book_2020}.
If the process of cluster landing on a surface significantly modifies their shape and morphology,
the major technical effort required to produce size-selected cluster beams is largely in vain.
Stability of size-selected clusters on a surface is therefore a key goal of deposition studies.

Stability and electronic properties of small clusters deposited on solid surfaces were also
studied theoretically within the framework of the liquid drop model
\cite{Poenary_2007_EPL.79.63001, Poenaru_2008_EPJD.47.379, Semenikhina_2008_JETP.106.678}.
Analytical relations were derived for the deformation-dependent surface and curvature energies of
small Na$_N$ and Ar$_N$ ($N < 150$) clusters of different shapes (oblate and prolate spheroids,
semi-spheroids and truncated ellipsoids) deposited on a solid surface.
Sequences of ``magic'' numbers for spheroidal and semi-spheroidal clusters were determined
for different values of the deformation parameter \cite{Poenaru_2008_EPJD.47.379}.
These results demonstrated the important role of deformation and quantum effects
in determining the stability of small atomic clusters on a substrate.

This paper reports a computational study of structure and stability of large (3 nm in diameter)
silver, gold and titanium clusters, Ag$_{887}$, Au$_{887}$ and Ti$_{787}$, deposited on graphite.
To the best of our knowledge, a comparative analysis of the dynamics of clusters made of different elements
and deposited at the same conditions has been lacking in the previous MD papers, each of which has focused
on one type of atomic clusters.
Other than that, an important question in MD simulations of cluster deposition concerns the choice of an
interaction potential between the cluster and the substrate.
Since the cluster--substrate interaction at soft-landing conditions is weak,
the widely accepted approach is to describe this interaction using pairwise Lennard-Jones or Morse potentials.
However, a literature review reveals that parameters of these potentials vary significantly across publications.
Therefore, it is an open question how the parameters of metal--surface interatomic interactions would affect
the stability of deposited clusters.

In this paper, we focused on the deposition energy range of $0.001 - 5.0$~eV per atom,
which corresponds to the soft-landing regime utilized in numerous experiments (e.g.,
Refs.~\cite{Brechignac_2002_PRL.88.196103, Lando_2007_EPJD.43.151, Solovyov_2014_PSSB.251.609,
Couillard_2003_APL.82.2595} to name a few).
Other phenomena that arise at more energetic collisions, e.g. cluster pinning or the formation of holes
due to penetration of clusters through a substrate \cite{Pratontep_2003_PRL.90.055503,
Smith_1993_ProcRoyalSocA.441.495, Plant_2014_Nanoscale.6.1258}, are beyond the scope of this study but
can also be simulated using the computational approach described below in Sect.~\ref{sec:methods}.

Parameters for the interaction between metal atoms and graphite are obtained from complementary
\textit{ab initio} calculations employing the second-order M{\o}ller-Plesset (MP2) perturbation theory.
We found that the shape of deposited clusters depends strongly on the element type.
The positions of cluster surface atoms obtained from MD simulations are fitted with a surface equation.
From this fit, the contact angle, contact radius and height of the clusters
are determined as functions of deposition energy.
The dependence of these parameters on the initial cluster structure and internal energy of the clusters is also analyzed.
Pre-equilibration of clusters at elevated temperatures results in a significant decrease
of the contact angle, although this trend is different for the silver, gold and titanium clusters.
The shape of Ti$_{787}$ is rather stable at cluster temperatures up to 900~K and deposition energies up to
0.25~eV/atom, whereas the shape of thermally excited Ag$_{887}$ and Au$_{887}$ clusters changes significantly
even at low deposition energies.

\section{Computational methodology}
\label{sec:methods}

The simulations were performed using MBN Explorer \cite{MBNExplorer_JCompChem_2012} -- a software package
for advanced multiscale modeling of complex molecular structure and dynamics.
MBN Studio \cite{MBNStudio_paper_2019}, a dedicated graphical user interface for MBN Explorer,
was used to construct the systems, prepare all necessary input files and analyze simulation outputs.

\subsection{Metal clusters}

As a first step, spherical clusters with the diameter of 3~nm were cut from ideal silver, gold and titanium crystals.
The resulting structures contained $N_{{\rm Ag}}$ = $N_{{\rm Au}}$ = 887 and $N_{{\rm Ti}}$ = 787 atoms.
Energy minimization calculations were conducted for the free clusters using the velocity quenching algorithm
\cite{MBNExplorer_UsersGuide} with the time step of 1~fs.
Interatomic interactions were described using the many-body Gupta potential~\cite{Gupta}
and the parameters were taken from Ref.~\cite{Cleri_1993_PRB.48.22}.

\begin{figure}[tb!]
\centering
\includegraphics[width=0.46\textwidth,clip]{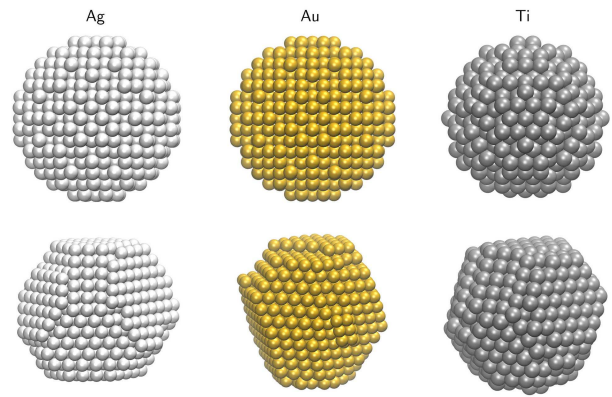}
\caption{Geometries of Ag$_{887}$, Au$_{887}$ and Ti$_{787}$ clusters obtained after energy minimization (upper row)
and the annealing process (bottom row).}
\label{fig:opt-vs-ann}
\end{figure}

Energy-minimized structures were annealed by means of MD simulations following
the computational protocol from Ref.~\cite{Ellaby_2018_JPCM.30.155301}.
The annealing procedure enabled sampling of the configuration space of the clusters at
elevated temperatures, at which the cluster surface has undergone a phase transition but the core remained,
at least partly, in the solid phase.
This protocol was validated \cite{Ellaby_2018_JPCM.30.155301} by the comparison of annealed cluster
geometries with the structures obtained from scanning transmission electron microscopy experiments.
The clusters were heated up to 700~K for Au$_{887}$, 800~K for Ag$_{887}$ and 900~K for Ti$_{787}$.
These values are about $70-100$~K lower than the melting temperatures of the clusters which were
determined through the analysis of caloric curves and root-mean-square displacement of all atoms.
Each cluster was heated from 0~K up to the target temperature for 1~ns, then kept at this temperature for 2~ns
and cooled down to 0~K over 1~ns, so that one annealing cycle was completed in 4~ns.
Three subsequent cycles were performed for each cluster to get energetically favorable structures.
Potential energy of each cluster decreased upon annealing by about $0.01 - 0.02$ eV/atom with respect
to the initial values obtained from energy minimization.
The annealed cluster structures contained several grains with different (mainly, fcc and hcp) crystal lattices.
Final structures of the energy-minimized and annealed clusters are compared in Fig.~\ref{fig:opt-vs-ann}.

\subsection{Graphite substrate}

The constructed graphite substrate contained 5 monolayers with the size of $106.35 \times 104.38$~\AA$^2$.
These dimensions were chosen to replicate the system with periodic boundary conditions.
The substrate size was more than three times larger than the cluster size and
significantly exceeded the range of interatomic interactions as described below.
Thus, we ensured that there is no artificial interaction across the simulation box boundaries.
Prior simulation of cluster deposition, the substrate was energy minimized using the velocity quenching algorithm.
The many-body Brenner potential \cite{Brenner_1990_PRB.42.9458} was used to describe the interaction
between covalently-bonded carbon atoms within each graphite layer whereas the Lennard-Jones potential
was employed to account for the van der Waals interaction between the layers.
The Lennard-Jones potential is implemented in MBN Explorer in the following form:
\begin{equation}
U(r) = \varepsilon \left[ \left( \frac{r_0}{r} \right)^{12}  -  2 \left( \frac{r_0}{r} \right)^{6}  \right] \ ,
\label{eq:LJ_r-min}
\end{equation}
where $\varepsilon$ is the depth of the potential energy well and $r_0$ is the equilibrium interatomic distance.
Parameters for the carbon--carbon interaction (see Table~\ref{table:LJ_parameters}) were taken from Ref.~\cite{Geng_2009_JPCC.113.6390}.
The Lennard-Jones potential was truncated at a cutoff distance of 10~\AA~that is about 3 times greater than
the interplanar distance in graphite and an order of magnitude smaller than the size of the simulation box.

\subsection{Metal--carbon interaction}

\begin{sloppypar}
The cluster--substrate interaction was also described by the Lennard-Jones potential using the
cutoff distance of 10~\AA.
Several MD studies of noble and transition-metal clusters (Ag, Au, Pt, Cu, Ni)
interacting with graphite and graphene were reported earlier, see e.g. \cite{Ryu_2010_JPCC.114.2022,
Luedtke_1999_PRL.82.3835, Neek-Amal_2009_Nanotechnology.20.135602}.
Parameters for the metal--carbon interaction were derived in these papers using empirical mixing rules:
\begin{equation}
r_0^{\textrm{C--M}} = \frac12 \left( r_0^{\textrm{C--C}} + r_0^{\textrm{M--M}} \right) \ \  , \ \
\varepsilon^{\textrm{C--M}} = \sqrt{ \varepsilon^{\textrm{C--C}} \, \,  \varepsilon^{\textrm{M--M}} } \ ,
\label{eq:mixing_rules}
\end{equation}
where $r_0^{\textrm{M--M}}$ and $\varepsilon^{\textrm{M--M}}$ are the parameters of
metal--metal interactions \cite{Erkoc_2001_AnnuRevCompPhys.IX.1}.
Other studies have shown \cite{Verkhovtsev_2014_EPJD.68.246, Galashev_2019_PhysLettA.383.252}
that parameters for the pairwise interactions between metal atoms and $sp^2$ carbon systems,
such as fullerenes or carbon nanotubes, differ significantly from those derived using the mixing rules.
Thus, a broad range of parameters for various metal--carbon systems can be found in literature,
and the optimal choice of the parameters is not obvious.
\end{sloppypar}

\begin{table}[tb!]
\centering
\caption{Parameters of the Lennard-Jones potential used in this work to describe the C--C interaction
between graphite layers as well as between the metal (Ag, Au, Ti) and carbon atoms.
Other values (of the Lennard-Jones or the pairwise Morse potentials) reported in literature are also
listed for comparison.}
\begin{tabular}{cp{0.3cm}p{1.6cm}p{1.4cm}p{2.7cm}}
\hline
  &  & $\varepsilon$~(eV) &  $r_0$~(\AA) &  Ref.  \\
\hline
 C--C      &   & 0.00286  &     3.89     & \cite{Geng_2009_JPCC.113.6390}                  \\
\hline
 Ag--C     &   &  0.020   &     3.49     &    this work (MP2)                              \\
           &   &  0.029   &     3.32     & \cite{Ryu_2010_JPCC.114.2022}                   \\
           &   &  0.030   &     3.37     & \cite{Neek-Amal_2009_Nanotechnology.20.135602}  \\
           &   &  0.009   &     3.45     & \cite{Galashev_2019_PhysLettA.383.252} (Morse)  \\
\hline
 Au--C     &   &  0.044   &     3.49     & this work (MP2)                                 \\
           &   &  0.033   &     3.32     & \cite{Ryu_2010_JPCC.114.2022}                   \\
           &   &  0.013   &     3.36     & \cite{Luedtke_1999_PRL.82.3835}                 \\
\hline
 Ti--C     &   &  0.165   &     2.44     & this work (MP2)                                 \\
\hline
 Ni--C     &   &  0.023   &     3.20     & \cite{Ryu_2010_JPCC.114.2022}                   \\
           &   &  0.345   &     2.03     & \cite{Verkhovtsev_2014_EPJD.68.246} (Morse)     \\
           &   &  0.363   &     2.28     & \cite{Galashev_2019_PhysLettA.383.252} (Morse)  \\
\hline
\end{tabular}
\label{table:LJ_parameters}
\end{table}

To elaborate on this issue, we performed {\it ab initio} calculations of
potential energy scans for Ag, Au and Ti atoms interacting with a benzene molecule,
which can be considered as a smallest structural unit of a graphite layer.
It is known (see, e.g., Refs.~\cite{Obolensky_2007_IntJQuantChem.107.1335, Grimme_2011_WIRE.1.211} and references therein)
that standard DFT methods do not account properly for long-range dispersion interactions
and require additional empirical corrections.
The importance of non-local correlation effects that govern dispersive interactions was particularly emphasized
for the binding of atoms and small clusters of silver and gold on graphite \cite{Amft_2011_JPCM.23.395001, Jalkanen_2007_JPCA.111.12317}.

Since the dispersive interaction is naturally accounted for in wave function-based \textit{ab initio} methods,
a series of calculations employing the second-order M{\o}ller-Plesset (MP2) perturbation theory were performed.
The Gaussian 09 software package \cite{g09} using a LanL2DZ basis set was employed.
The metal atoms were placed in the hollow position on top of the benzene molecule and displaced along its main axis.
The metal atom--benzene interaction energy was obtained from these scans
and divided by the number of carbon atoms to determine the interaction energy per M--C pair of atoms.
Since benzene is a highly symmetric molecule and the metal atoms were displaced along its main axis,
all M--C bonds can be considered as equivalent.
The evaluated parameters of the Lennard-Jones potential for Ag--C, Au--C and Ti--C are summarized
in Table~\ref{table:LJ_parameters}.
Other values reported in literature \cite{Luedtke_1999_PRL.82.3835, Ryu_2010_JPCC.114.2022,
Neek-Amal_2009_Nanotechnology.20.135602, Galashev_2019_PhysLettA.383.252} are also presented for completeness.
Note that the parameters from Refs.~\cite{Ryu_2010_JPCC.114.2022, Luedtke_1999_PRL.82.3835,
Neek-Amal_2009_Nanotechnology.20.135602} were determined using the mixing rules (Eq.~(\ref{eq:mixing_rules}))
while the parameters from Refs.~\cite{Verkhovtsev_2014_EPJD.68.246, Galashev_2019_PhysLettA.383.252}
were derived on the basis of DFT calculations.
To the best of our knowledge, there is no reference data available on the interaction between
Ti and C atoms.
Therefore, in Table~\ref{table:LJ_parameters} we provide also parameters taken from literature
\cite{Ryu_2010_JPCC.114.2022, Verkhovtsev_2014_EPJD.68.246, Galashev_2019_PhysLettA.383.252}
for the interaction between carbon and nickel, another open-shell transition metal.

Our MP2 calculations revealed that the metal--carbon interaction energy increases from silver to gold to titanium.
The Au--C potential well depth is about two times larger than for the Ag--C interaction.
This is in agreement with the results of earlier DFT calculations with dispersion corrections
\cite{Amft_2011_JPCM.23.395001} which found that adsorption energy of a gold atom on graphite
is about two times higher than for a silver atom.
It was also found~\cite{Amft_2011_JPCM.23.395001} that no charge redistribution occurs between
the Ag atom and graphite, while the deposited Au atom receives a charge of approximately 0.1$e$
from the carbon sheet.
On this basis, it was concluded that the adsorption of silver on graphite is purely of van der Waals type
whereas small hybridization in the density of states, i.e. a chemical contribution to the binding,
occurs in the case of gold adsorbed on graphite.
As it is indicated in Table~\ref{table:LJ_parameters}, the Ti--C interaction is about eight times stronger
than the Ag--C interaction.
A recent DFT-based study of the titanium/graphite interface \cite{Chen_2020_JPCM.32.145001} reported
the formation of chemical bonding between interfacial Ti and C atoms.
It was found also that an interfacial Ti atom acquires a charge of about $0.3e$.
Another DFT study of the adsorption of a titanium slab on graphene demonstrated that a $p-d$
hybridization occurs between atomic orbitals of carbon and titanium \cite{Hsu_2014_ACSNano.8.7704}.


\subsection{Deposition of clusters on graphite}

\begin{sloppypar}
For MD simulations of cluster deposition, a simulation box of $106.35 \times 104.38 \times 110$~\AA$^3$ was used.
Each cluster was placed in the center of the simulation box approximately 40~\AA~above the topmost graphite layer.
The clusters were deposited with energies $E_{{\rm dep}} = 0.001, 0.01, 0.05, 0.1, 0.25, 0.75$~and 1.0~eV/atom
at normal incidence to the graphite surface.
To determine fragmentation threshold for the Ti$_{787}$ cluster, higher deposition energies
from 2.0 to 5.0~eV/atom were also considered.
\end{sloppypar}

To simulate the deposition of clusters and their rearrangement on the surface,
250-ps long MD simulations were performed for the microcanonical ($NVE$) ensemble of particles.
The simulation time was chosen such that, after hitting the surface, the clusters would relax for at least 200~ps.
Kinetic and potential energies of the system reached steady-state values within $20-40$~ps
after the collision and remained practically constant until the end of each simulation.
Integration of equations of motion was done using the velocity Verlet algorithm with the time step of 1~fs.
We ensured that a variation of the total energy of the system did not exceed 0.01\% with this time step.
Two bottom graphite layers were fixed to avoid translational motion of the whole system after the collision.
Test calculations were performed also for thicker graphite substrates containing 7 and 11 monolayers
to check that the results obtained do not depend on substrate thickness.

As described in Section~\ref{sec:results},
disintegration of the clusters and scattering of cluster fragments over the whole graphite sample
were observed at deposition energies of 0.75 eV/atom and above.
To ensure that the simulation box boundaries do not affect the results, a set of simulations
was performed on a larger graphite substrate with the size of $212.7 \times 208.8$~\AA$^2$. 
The total number of atoms in the systems thus varied from approx. 19,000 to 86,000.

\section{Results and discussion}
\label{sec:results}

We begin our analysis by considering deposition of the spherical Ag$_{887}$, Au$_{887}$ and Ti$_{787}$ clusters
cut from ideal bulk crystals.
Then we discuss how alteration of the initial cluster structure due to annealing affects their shape upon deposition.
Finally, we consider deposition of thermally excited clusters and compare the resulting shapes
with the case of deposition at zero temperature.

\begin{sloppypar}
For cluster sizes smaller than we considered in this study, quantum effects such as even-odd oscillations
in cluster abundance spectra \cite{Katakuse_1985_IJMS.67.229} and the appearance of ``magic'' numbers
associated with electron shell closure \cite{Haekkinen_2016_AdvPhysX.1.467}, become more prominent and
play a crucial role in determining the shape of clusters on a surface \cite{Poenaru_2008_EPJD.47.379}.
Such effects are particularly strong for the clusters containing $N \lesssim 200$ atoms but shrink with increasing
the cluster size up to $N \approx 800$ \cite{Reinhard_ClusterDynamics}.
\end{sloppypar}

\begin{figure*}[htb!]
\centering
\includegraphics[width=0.98\textwidth]{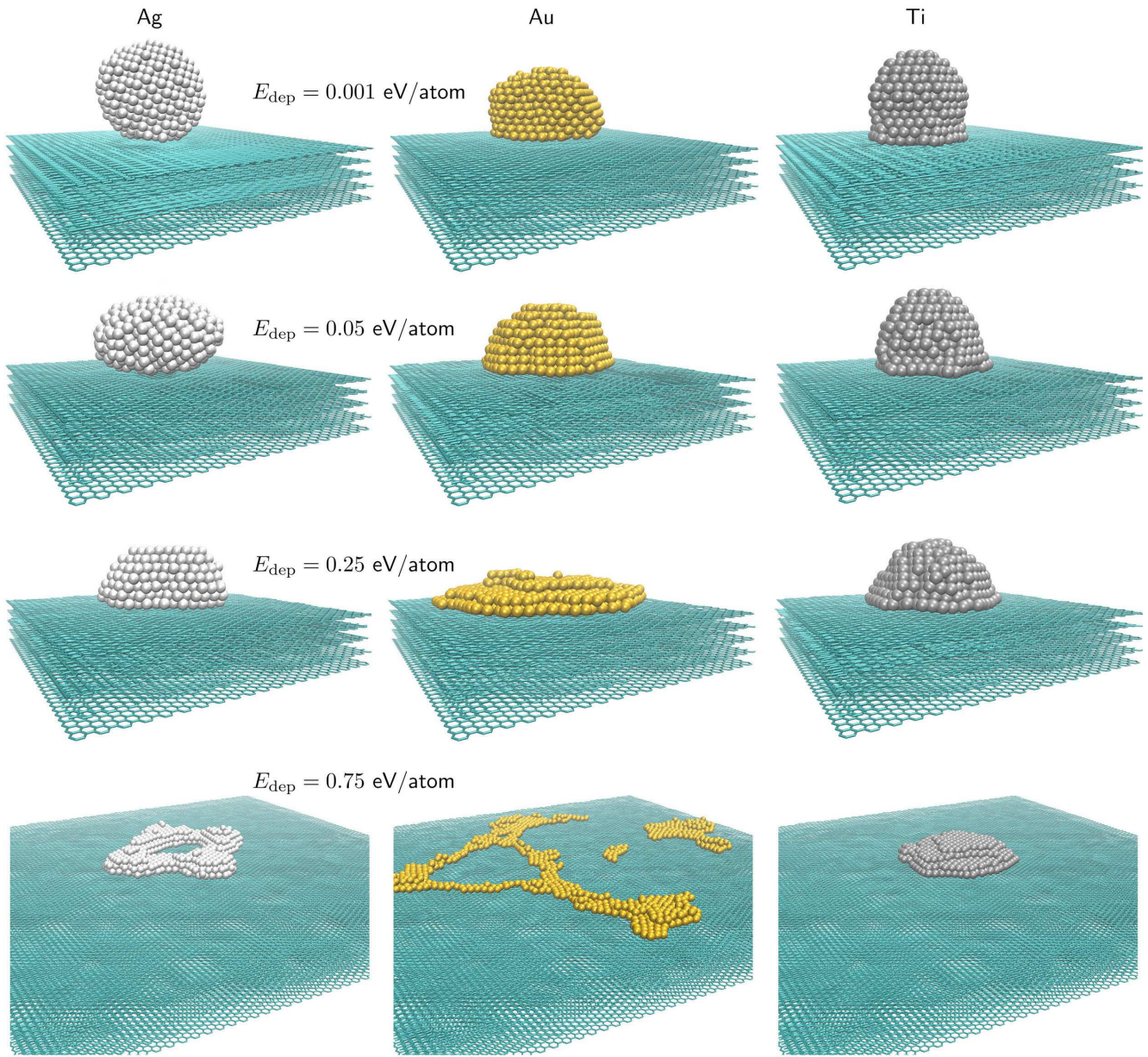}
\caption{MD snapshots of the spherical Ag$_{887}$ (left column), Au$_{887}$ (middle column) and Ti$_{787}$ (right column)
clusters deposited on graphite at the deposition energies of 0.001, 0.05, 0.25 and 0.75~eV/atom (top to bottom rows).}
\label{fig:cluster_snapshots}
\end{figure*}

\begin{sloppypar}
Figure~\ref{fig:cluster_snapshots} shows final snapshots of the spherical Ag$_{887}$, Au$_{887}$ and Ti$_{787}$ clusters
deposited on graphite at different deposition energies $E_{{\rm dep}}$.
The figure indicates clearly that the cluster geometry depends on the element type
and that it changes significantly with an increase of deposition energy.
At $E_{{\rm dep}} = 0.001$~eV/atom (upper row) the Ag$_{887}$ cluster retains its spherical shape,
whereas the gold and titanium clusters are deformed due to stronger adhesion to the surface.
The Au$_{887}$ cluster has the shape of a truncated ellipsoid while Ti$_{787}$ is a capped structure
elongated in the direction normal to the substrate.
At $E_{{\rm dep}} = 0.05$~eV/atom (second row) the silver cluster acquires a slightly deformed
quasi-ellipsoidal shape, while both gold and titanium clusters transform into truncated ellipsoids.
All three clusters deposited at $E_{{\rm dep}} = 0.25$~eV/atom (third row) transform into truncated ellipsoids
with the gold cluster being the most flattened structure.
Finally, at $E_{{\rm dep}} = 0.75$~eV/atom (bottom row) topology of the silver and gold clusters
changes after collision with the surface.
The Ag$_{887}$ cluster becomes a hollow structure that remains stable over the 250-ps long simulation.
The Au$_{887}$ cluster fragments into a large pretzel-like structure and several smaller islands that
are scattered over the surface.
In contrast, Ti$_{787}$ remains intact but gradually becomes more and more flattened.
Note that the simulations performed at $E_{{\rm dep}} = 0.75$~eV/atom were conducted on a large
graphite substrate of $212.7 \times 208.8$~\AA$^2$.
\end{sloppypar}

\begin{figure*}[htb!]
\centering
\includegraphics[width=0.96\textwidth]{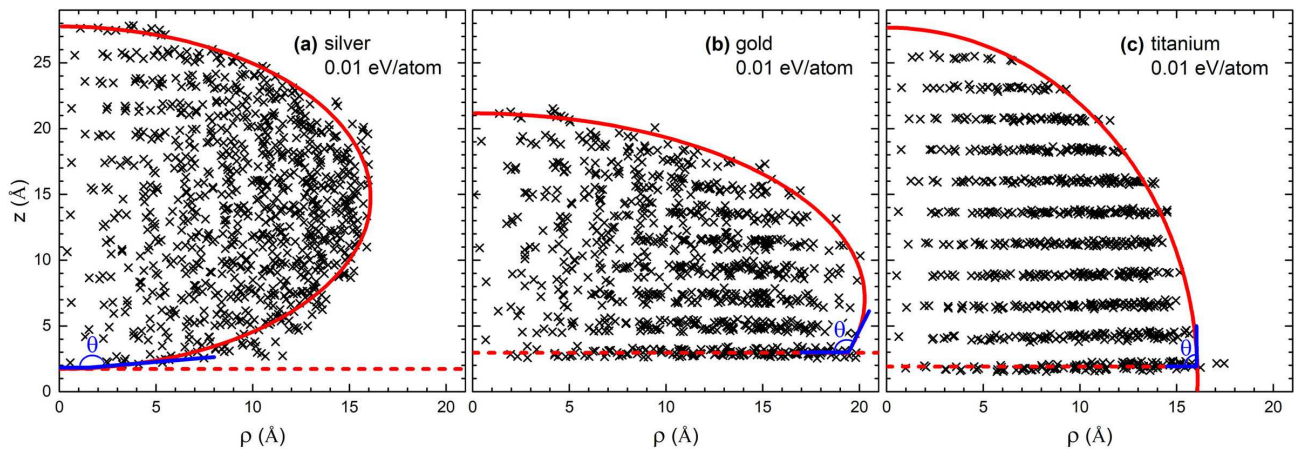}
\caption{
Radial profiles of the initially spherical Ag$_{887}$ (a), Au$_{887}$ (b) and Ti$_{787}$ (c) clusters deposited
at $E_{{\rm dep}} = 0.01$~eV/atom. 
Symbols (crosses) show the distribution of all atoms in each cluster at the end of a 250~ps-long simulation.
Solid curves show the best fit of the cluster profile with $z(\rho)$ that is an inverse function of $\rho(z)$, Eq.~(\ref{eq:surface_equation}).
Dashed lines indicate the averaged position of the bottom-most atomic layer in each cluster.}
\label{fig:cluster_shape}
\end{figure*}

Coordinates of cluster atoms, extracted from each MD trajectory, were used to parameterize
the cluster shape and to evaluate a contact angle with the substrate.
As follows from the simulated trajectories, each deposited cluster is, to a good approximation,
radially symmetric with respect to its main axis.
Thus, we introduced cylindrical coordinates $\rho$ and $z$, where
\begin{equation}
\rho = \sqrt{(x - x_{\rm CM})^2 + (y - y_{\rm CM})^2 } \ ,
\end{equation}
with $x_{\rm CM}$ and $y_{\rm CM}$ being $x$- and $y$-projections of the center of mass of each cluster.
The $\rho$-axis lies in the graphite surface plane, whereas $z$-axis is perpendicular
to the surface and $z=0$ corresponds to the average position of the topmost graphite layer.
Figure~\ref{fig:cluster_shape} shows by symbols ($\rho, z$) projections of all atoms
in the Ag$_{887}$, Au$_{887}$ and Ti$_{787}$ clusters deposited at 0.01~eV/atom.
Figure~\ref{fig:cluster_shape}(a) shows that at low values of $E_{{\rm dep}}$ the Ag$_{887}$ cluster acquires
the shape of a non-truncated ellipsoid which is slightly deformed from the side which is in contact with graphite.
The initially spherical Au$_{887}$ cluster transforms into a truncated oblate spheroid (Fig.~\ref{fig:cluster_shape}(b))
while the Ti$_{787}$ cluster has a well pronounced prolate shape, i.e. its height is larger than the contact radius (Fig.~\ref{fig:cluster_shape}(c)).
Interestingly, atoms in the titanium cluster arrange into layers oriented parallel to graphite planes,
that is different from the atomic arrangement in the silver and gold clusters.
This can be explained by a stronger interaction between titanium and carbon atoms as compared to the Ag--C and
Au--C interactions.

For each cluster we selected coordinates of atoms located on the surface and fitted the resulting profiles
with the following surface equation \cite{Giovambattista_2007_JPCB.111.9581, Skvara_2018_MolSimul.44.190}:
\begin{equation}
\rho(z) = \left\{
\begin{array}{l l}
\sqrt{ a(z - z_0)^2 + b(z - z_0) + c }   &  , \ \ z \ge 0 \
\\
0           &  , \ \ z < 0 \
\end{array} \right. ,
\label{eq:surface_equation}
\end{equation}
where $a$, $b$, $c$ and $z_0$ are fitting parameters.
This expression enables a description of different cluster shapes with a single fitting function
without any geometrical assumptions on the cluster shape.
From the least-squares fit of cluster profiles with Eq.~(\ref{eq:surface_equation})
we determined the contact radius and height of each cluster as well as the contact angle
with the substrate as functions of deposited energy.
An inverse dependence $z(\rho)$ is shown by solid red curves in Fig.~\ref{fig:cluster_shape}.

The contact angle $\theta$ was evaluated by calculating the derivative of $z(\rho)$ at the point $\rho^{\prime}$
which corresponds to the average position of the bottom-most atomic layer of the clusters, $z^{\prime} = z(\rho^{\prime})$,
see the dashed lines in Fig.~\ref{fig:cluster_shape}.
An expression for the contact angle is then given by \cite{Giovambattista_2007_JPCB.111.9581}
\begin{equation}
\theta
= \arctan \left( \left. \frac{dz}{d\rho} \right|_{\rho = \rho^{\prime}} \right)
= \frac{\pi}{2} + \arctan \left( \left. \frac{d\rho}{dz} \right|_{z = z^{\prime}} \right) \ .
\label{eq:contact_angle}
\end{equation}

\begin{figure}[htb!]
\centering
\includegraphics[width=0.42\textwidth]{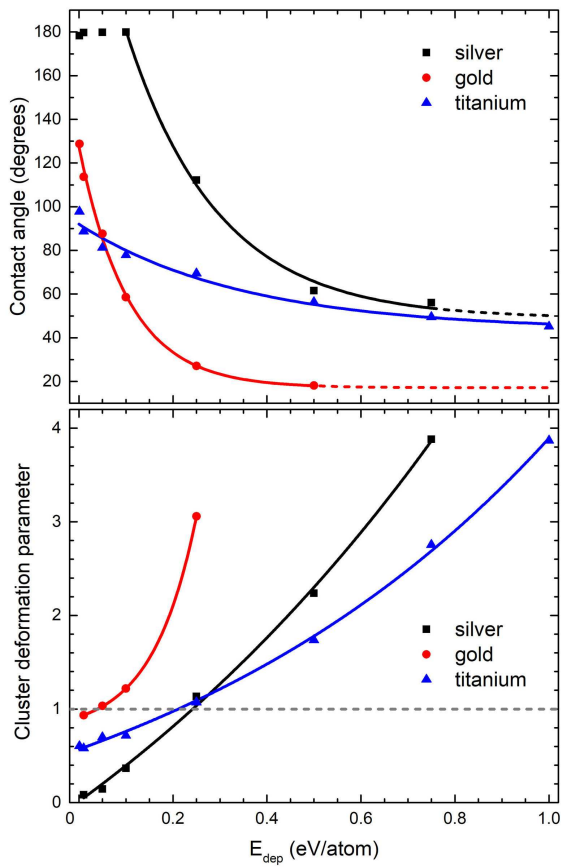}
\caption{
Dependence of the contact angle (Eq.~(\ref{eq:contact_angle}), top panel) and
cluster deformation parameter (Eq.~(\ref{eq:aspect_ratio}), bottom panel) on deposition energy
for the initially spherical Ag$_{887}$, Au$_{887}$ and Ti$_{787}$ clusters.
Solid curves show exponential fits using Eqs.~(\ref{eq:angle_exp_fit}) and (\ref{eq:ratio_exp_fit}). }
\label{fig:contact_angle}
\end{figure}

\begin{figure*}[htb!]
\centering
\includegraphics[width=0.96\textwidth]{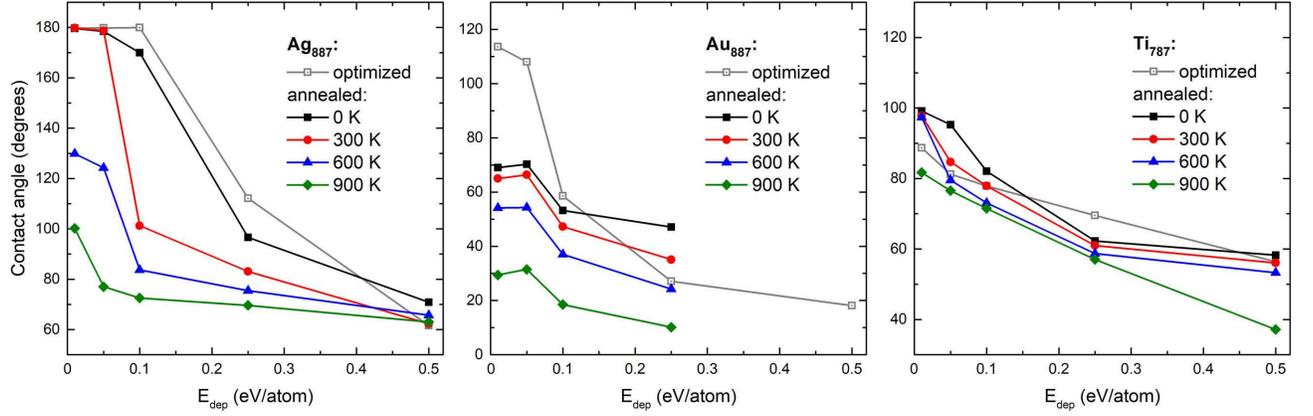}
\caption{Contact angle $\theta$ as a function of deposition energy for the Ag$_{887}$, Au$_{887}$ and Ti$_{787}$ clusters,
either optimized (open symbols) or annealed and then equilibrated at a given temperature (closed symbols).}
\label{fig:contact_angle_temperature}
\end{figure*}

\begin{figure*}[htb!]
\centering
\includegraphics[width=0.96\textwidth]{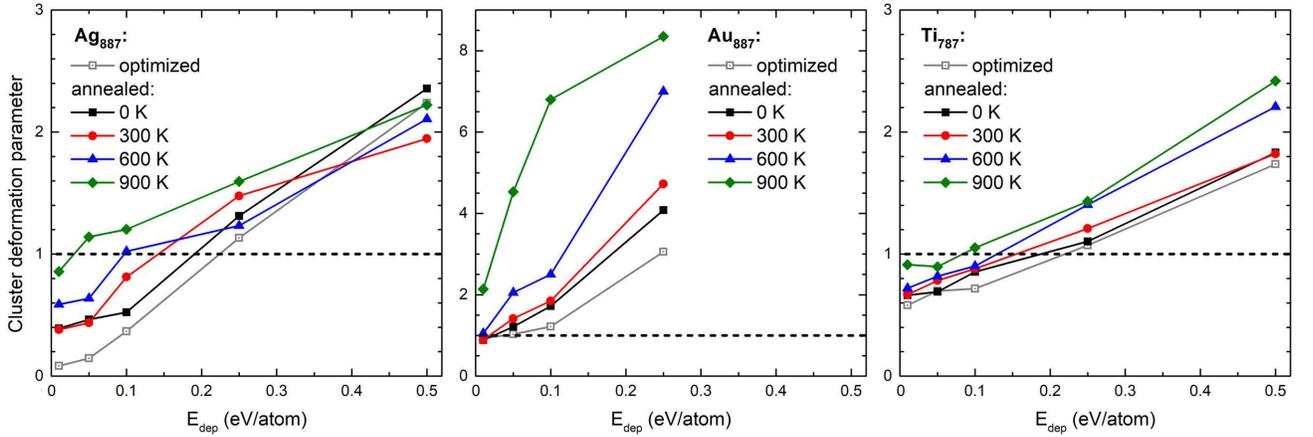}
\caption{
The cluster deformation parameter $\delta$ as a function of deposition energy for the Ag$_{887}$, Au$_{887}$ and Ti$_{787}$ clusters,
either optimized (open symbols) or annealed and then equilibrated at a given temperature (closed symbols).}
\label{fig:cluster_shape_temperature}
\end{figure*}

The contact angle for the optimized Ag$_{887}$, Au$_{887}$ and Ti$_{787}$ clusters as a function of
deposition energy is presented in the upper panel of Fig.~\ref{fig:contact_angle}.
The calculated values of $\theta$ were averaged over three distinct cluster geometries sampled from the
last 100~ps of each simulation.
It was found that the deposited clusters acquire their equilibrium shapes shortly after reaching the surface
(on the timescales of few tens of picoseconds) and a variation of $\theta$ in the remaining part of simulations
does not exceed 10 degrees.

\begin{sloppypar}
The figure shows that the contact angle for the Ag$_{887}$ cluster (black squares) evolves differently with
an increase of $E_{{\rm dep}}$ than for the gold and titanium clusters.
The main distinction is that $\theta$ for Ag$_{887}$ remains close to 180$^{\circ}$ at deposition energies up to 0.1~eV/atom.
A further increase of $E_{{\rm dep}}$ up to 0.25~eV/atom leads to a rapid decrease of the angle by about 70 degrees whereas
it converges to a value of about 60$^{\circ}$ at $E_{{\rm dep}} = 0.75$~eV/atom.
At higher deposition energies, the cluster fragments into several small islands
and therefore the contact angle was not determined.
A similar trend was observed for Au$_{887}$ (red circles).
In this case, the contact angle drops rapidly from approx.~130$^{\circ}$ at $E_{{\rm dep}} = 0.001$~eV/atom
down to 20$^{\circ}$ at $E_{{\rm dep}} = 0.5$~eV/atom.
In the latter case, the cluster is rearranged into two atomic layers which are distributed uniformly
over large graphite area.
Deposition at higher energies also leads to cluster fragmentation and the formation of smaller islands
with the height of two atomic layers.
For the Ti$_{787}$ cluster (blue triangles), $\theta$ gradually decreases from 100$^{\circ}$ to 45$^{\circ}$ in
the deposition energy range considered.
This corresponds to the observations shown in Fig.~\ref{fig:cluster_snapshots} that the
contact area for Ti$_{787}$ increases gradually with $E_{{\rm dep}}$.
Interestingly, the contact angle for all three clusters considered decreases exponentially with $E_{{\rm dep}}$,
\begin{equation}
\theta = \theta_0 + \theta_1 \, e^{ -\alpha E_{{\rm dep}} } \ ,
\label{eq:angle_exp_fit}
\end{equation}
see solid curves in the upper panel of Fig.~\ref{fig:contact_angle}.
The corresponding fitting parameters are summarized in Table~\ref{table:angle_fitting}.
\end{sloppypar}

\begin{table}[tb!]
\centering
\caption{Fitting parameters describing an exponential decrease of the contact angle (Eq.~(\ref{eq:angle_exp_fit}))
and an increase of the cluster deformation parameter (Eq.~(\ref{eq:ratio_exp_fit})) with $E_{{\rm dep}}$.
Note that only the data points for $E_{{\rm dep}} \ge 0.1$~eV/atom were fitted with Eq.~(\ref{eq:angle_exp_fit})
for the Ag$_{887}$ cluster.}
\begin{tabular}{p{2.4cm}cp{0.2cm}cp{0.2cm}c}
\hline
                            &  Ag$_{887}$ & &  Au$_{887}$ & &  Ti$_{787}$  \\
\hline
 $\theta_0$~(deg.)          &    48.87    & &     17.11   & &   43.47      \\
 $\theta_1$~(deg.)          &   219.93    & &    110.30   & &   48.52      \\
 $\alpha$~(eV$^{-1}$/atom)  &     5.13    & &      9.58   & &    2.83      \\
\hline
 $\delta_0$                 &    -4.29    & &      0.76   & &   -1.04      \\
 $\delta_1$                 &     4.30    & &      0.16   & &    1.61      \\
 $\beta$~(eV$^{-1}$/atom)   &     0.85    & &     10.74   & &    1.12      \\
\hline
\end{tabular}
\label{table:angle_fitting}
\end{table}

Additional MD simulations were conducted at higher deposition energies to determine
the fragmentation threshold for Ti$_{787}$.
We found that the threshold deposition energy is between 2 and 3 eV per atom,
which is about four times larger than the corresponding fragmentation thresholds for Ag$_{887}$ and Au$_{887}$.
At $E_{{\rm dep}} = 2$ eV/atom the titanium cluster is flattened over the graphite surface but remains intact,
whereas at $E_{{\rm dep}} = 3$ eV/atom the cluster transforms into a flat pretzel-like structure
(similar to the Au$_{887}$ cluster deposited at 0.75~eV/atom, see Fig.~\ref{fig:cluster_snapshots})
with the maximal height of two atomic layers.
Deposition of Ti$_{787}$ at 4~eV/atom results in the formation of small titanium islands
that are scattered over the whole simulated substrate.
Note that the fragmentation thresholds should depend not only on elemental composition of
the clusters but also on their size.
However, we leave a detailed analysis of this dependence for further studies.

To complement this analysis, the lower panel of Fig.~\ref{fig:contact_angle} shows
the cluster deformation parameter $\delta$, defined as a ratio of  cluster contact radius
(radius of the bottom-most atomic layer, see Fig.~\ref{fig:cluster_shape}) to cluster height,
\begin{equation}
\delta = \frac{  \rho(z = z^{\prime})  }{  z(\rho = 0)  } \ ,
\label{eq:aspect_ratio}
\end{equation}
for the initially spherical Ag$_{887}$, Au$_{887}$ and Ti$_{787}$ clusters deposited at different energies.
The dependence of $\delta$ on $E_{{\rm dep}}$ also follows an exponential law,
\begin{equation}
\delta = \delta_0 + \delta_1 \, e^{ \beta E_{{\rm dep}} } \ ,
\label{eq:ratio_exp_fit}
\end{equation}
see solid curves in the lower panel of Fig.~\ref{fig:contact_angle}.
The corresponding fitting parameters are also listed in Table~\ref{table:angle_fitting}.
The horizontal dashed line denotes the case when the contact radius is equal to the height,
which corresponds to a perfect semi-spheroidal shape.
The figure shows that the shape of Au$_{887}$ is close to a semi-spheroid at low deposition energies (up to 0.05~eV/atom)
while highly oblate truncated spheroids are formed in the course of deposition at higher energies.
A similar trend is observed for the silver and titanium clusters but in this case the cluster shape is
close to a semi-spheroid at $E_{{\rm dep}} = 0.25$~eV/atom.
Thus, by increasing the deposition energy the clusters evolve from an ellipsoid (in the case of silver)
or a truncated prolate spheroid (in the case of titanium) to semi-spheroids to very flat structures with
the contact radius exceeding the cluster height by the factor of four.


The above described analysis was carried out for the spherical clusters deposited at zero temperature.
It is worth exploring how the shape of deposited clusters would change upon altering the
initial cluster structure due to annealing.
Results of this analysis are summarized in Figs.~\ref{fig:contact_angle_temperature} and \ref{fig:cluster_shape_temperature},
which show how the contact angle $\theta$ and the deformation parameter $\delta$ evolve as functions of $E_{{\rm dep}}$.
Gray lines / open symbols correspond to the spherical cluster case described above in Fig.~\ref{fig:contact_angle}.
Black lines / filled squared represent the results for the annealed clusters (see Fig.~\ref{fig:opt-vs-ann}) at
initial temperature of 0~K.
The figures show that alteration of the cluster structure due to annealing have a moderate impact on
the shape of Ag$_{887}$ and Ti$_{787}$ clusters in the deposition energy range considered,
whereas annealing of Au$_{887}$ results in a very different geometry even at low deposition energy.
At $E_{{\rm dep}} = 0.01$~eV/atom the annealed gold cluster wets the surface stronger than its spherical counterpart
and the contact angle decreases from 130$^{\circ}$ down to 70$^{\circ}$.
However, the variation of $\theta$ with $E_{{\rm dep}}$ is much smaller than for the spherical Au$_{887}$ cluster so
that at $E_{{\rm dep}} = 0.25$~eV/atom the contact angle for the annealed Au$_{887}$ is almost two times
larger than for the spherical cluster.
Deposition of the annealed cluster at higher energies results in its fragmentation.
Figure~\ref{fig:cluster_shape_temperature} indicates that the annealed noble metal clusters are characterized
by a larger deformation parameter as compared to the spherical clusters but annealing of the titanium Ti$_{787}$
cluster has a minor impact on its shape after deposition.

\begin{figure}[htb!]
\centering
\includegraphics[width=0.42\textwidth]{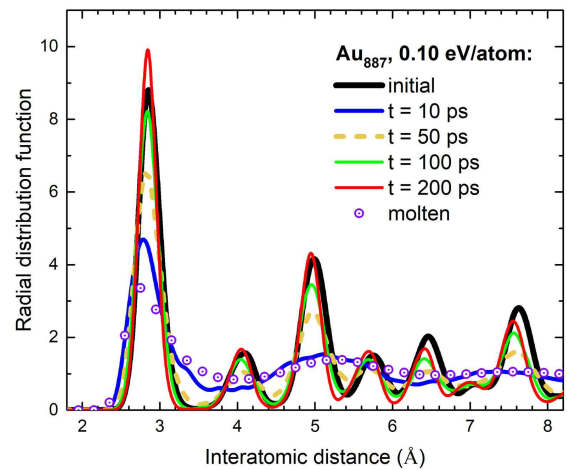}
\caption{Radial distribution function for the Au$_{887}$ cluster, pre-equilibrated at 300~K,
at different time instances $t$ after the collision.
An RDF for the molten state was obtained from a simulation of melting of a free Au$_{887}$ cluster.}
\label{fig:RDF_time}
\end{figure}

Finally, let us analyze how pre-equilibration of the clusters at different temperatures affects their
stability and shape upon deposition.
Figures~\ref{fig:contact_angle_temperature} and \ref{fig:cluster_shape_temperature} show
the contact angle $\theta$ (Eq.~(\ref{eq:contact_angle})) and the deformation parameter $\delta$ (Eq.~(\ref{eq:aspect_ratio}))
of the clusters after they were given an initial temperature of 300~K, 600~K and 900~K.
The latter value is just above the melting point of Ag$_{887}$ and Au$_{887}$ but about 100 degrees less than
the melting temperature of Ti$_{787}$.
Therefore, the silver and gold clusters equilibrated at 900~K are deposited as liquid droplets while the titanium
cluster has a molten surface but its core is still, at least partly, in the solid phase.
As one may expect, deposition of the clusters at elevated temperature leads to a significant decrease of the contact
angle and an increased contact area at low deposition energies.
The contact angle for Ag$_{887}$ saturates with an increase of $E_{{\rm dep}}$ at the value of about 60 degrees,
independent on the cluster initial temperature.
In contrast, the shape of Au$_{887}$ depends strongly on the amount of internal energy stored in the cluster.
While the profile of the $\theta(E_{{\rm dep}})$ dependence does not change when the initial temperature
is increased from 0~K to 900~K, a decrease of the contact angle is evident.

\begin{sloppypar}
The contact angle for the thermalized silver cluster deposited at energies below 0.1 eV/atom
exceeds the corresponding values for the gold and titanium clusters.
This result agrees with experimental observations that nanometer-size silver clusters deposited on graphite at
$E_{{\rm dep}} = 0.05$~eV/atom \cite{Brechignac_2002_PRL.88.196103, Lando_2007_EPJD.43.151, Solovyov_2014_PSSB.251.609}
are highly mobile, which leads to the formation of fractal-like silver nanostructures.
\end{sloppypar}

Further insights into the stability of clusters on the surface can be drawn from the analysis of
radial distribution function (RDF).
Figure~\ref{fig:RDF_time} shows an exemplary RDF for the Au$_{887}$ cluster, pre-equilibrated
at 300~K prior deposition, at different time instances $t$ after the collision.
Mechanical stress induced by the collision causes the formation of a liquid droplet
within the first 10~ps after the collision (solid blue curve).
Over the next 40~ps the droplet re-crystallizes into a solid structure whose RDF resembles the initial one (dashed yellow curve).
This structure remains stable and practically does not change in the remaining part of the simulation.

\begin{figure*}[htb!]
\centering
\includegraphics[width=0.94\textwidth]{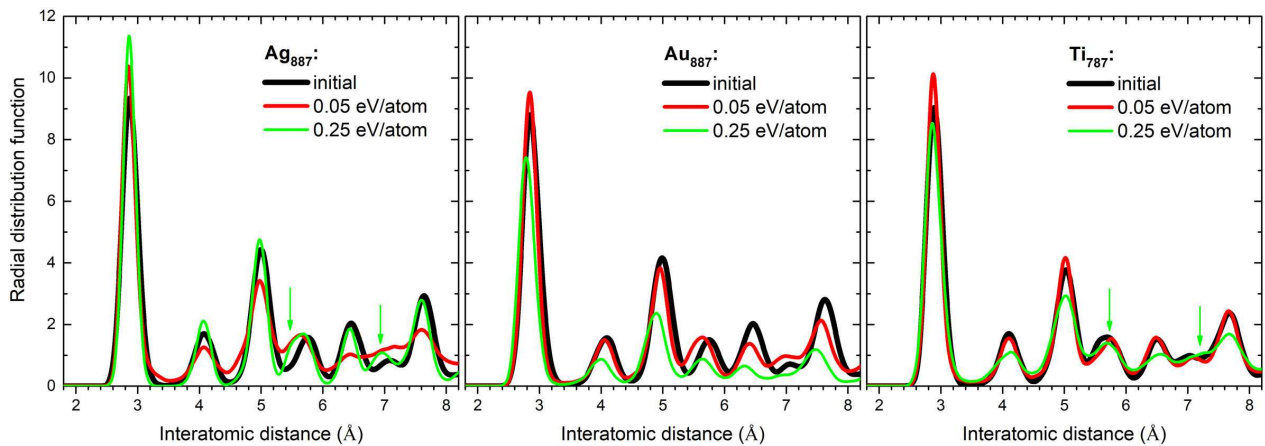}
\caption{
Radial distribution function for the Ag$_{887}$, Au$_{887}$ and Ti$_{787}$ clusters, pre-equilibrated at 300~K,
deposited at energies of 0.05 and 0.25 eV/atom.
Arrows show the appearance of fine structures in the RDFs that are indicative for
fcc $\to$ hcp (in the case of silver) and hcp $\to$ fcc (in the case of titanium) structural transformations. }
\label{fig:RDF_comparison}
\end{figure*}

Further analysis reveals that the clusters undergo structural transformations induced by the collision.
Figure~\ref{fig:RDF_comparison} compares RDFs for the annealed Ag$_{887}$, Au$_{887}$  and Ti$_{787}$ clusters
deposited at 0.05 and 0.25 eV/atom. The RDFs are calculated at the end of each 250-ps long trajectory.
Lattice structure of the gold cluster deposited at 0.25 eV/atom becomes compressed
as compared to the initial structure (see middle panel).
This is evident from that the fact that all the peaks in the RDF are strongly suppressed and uniformly
shifted towards smaller interatomic distances.
The silver cluster deposited at both 0.05 and 0.25~eV/atom maintains the short-range order (see left panel).
However, at $E_{{\rm dep}} = 0.05$~eV/atom the long-range order is lost as the peaks in the
interatomic distance range of $6-8$~\AA~are merged into a broader and more uniform distribution.
In this case, the cluster does not melt upon collision and then recrystallize (as it is shown in Fig.~\ref{fig:RDF_time})
but forms an amorphous-like structure without long-range order, which is stable on the simulation timescale of 250~ps.
Amorphization of the Ag$_{887}$ cluster deposited at 0.05~eV/atom can also be seen in Fig.~\ref{fig:cluster_snapshots}.
In contrast, lattice structure of the silver cluster deposited at $E_{{\rm dep}} = 0.25$~eV/atom resembles,
to a large extent, the initial lattice structure.
The only important difference is the formation of a shoulder at the interatomic distance of about 5.5~\AA~and
a shift of another peak centered at 7.1~\AA~to 6.9~\AA~(see green arrows).
This is an indication of an increased ratio of hcp lattice packing in the deposited cluster structure.
RDFs for the Ti$_{787}$ cluster (right panel) illustrate the opposite phenomenon, namely a shift of the peaks centered at 5.5~\AA~and 6.9~\AA~towards larger interatomic distances, that is indicative for an increased ratio of fcc lattice in the final
cluster structure.

\section{Conclusion}

The deposition of metal clusters made of three different elements -- silver, gold and titanium --
on a graphite substrate was studied by means of molecular dynamics simulations using the MBN Explorer
and MBN Studio software packages.
The clusters had the diameter of 3~nm and contained $N_{{\rm Ag}}$ = $N_{{\rm Au}}$ = 887 and $N_{{\rm Ti}}$ = 787 atoms.
We focused on deposition energies in the range $0.001 - 5.0$~eV/atom,
which corresponds to the soft-landing regime utilized in experiments.
Parameters for the interaction between metal atoms and a carbon surface were determined from
\textit{ab initio} calculations employing the second-order M{\o}ller-Plesset (MP2) perturbation theory.

We found that the shape and stability of deposited clusters depends strongly on the element type.
At low deposition energies, the Ag$_{887}$ cluster has a quasi-ellipsoid shape while
the gold and titanium clusters rearrange upon collision into truncated oblate and prolate spheroids.
Both silver and gold clusters flatten over the surface and eventually disintegrate as the deposition energy
increases up to $0.75 -1.0$~eV/atom, while the titanium cluster fragments at about four times higher energy.

The positions of cluster surface atoms obtained from MD simulations were fitted with a surface equation.
From this fit, the contact angle, contact radius and height of the clusters were determined as functions
of deposition energy.
We found that the contact angle for the noble metal and titanium clusters evolves differently with
an increase of deposition energy.
The contact angle for the silver cluster does not vary significantly at low deposition energies
up to 0.1~eV/atom, while it rapidly decreases by about 70 degrees at higher energies.
For the gold cluster, the angle drops rapidly from 130$^{\circ}$ down to 20$^{\circ}$ at $E_{{\rm dep}} = 0.5$~eV/atom.
The contact angle for the titanium cluster decreases gradually from 100$^{\circ}$ to 45$^{\circ}$ in the
range of deposition energies considered.

We also analyzed how the initial structure (optimized vs. annealed geometries)
and internal energy of the clusters affects the shape and stability of the deposited clusters.
Pre-equilibration of clusters at elevated temperatures up to 900~K results in a significant decrease
of the contact angle, although this trend is different for the silver, gold and titanium clusters.
The shape of Ti$_{787}$ is rather stable at cluster temperatures up to 900~K and at deposition energies up to
0.25~eV/atom, whereas the shape of thermally excited Ag$_{887}$ and Au$_{887}$ clusters changes significantly
even at low deposition energies.

\section*{Acknowledgements}

\begin{sloppypar}
This work was supported in part by Deutsche Forschungsgemeinschaft (Project no. 415716638)
and by the European Union's Horizon 2020 research and innovation programme
(the Radio-NP project within the H2020-MSCA-IF-2017 call, GA 794733 and
the RADON project within the H2020-MSCA-RISE-2019 call, GA 872494).
The possibility to perform calculations at
the Goethe-HLR cluster of the Frankfurt Center for Scientific Computing and at
the DeiC National HPC Center (University of Southern Denmark, Odense)
is gratefully acknowledged.
\end{sloppypar}





\end{document}